\documentclass[%
 aip,
 amsmath,amssymb,
 reprint,%
]{revtex4-1}

\usepackage{graphicx}
\usepackage{dcolumn}
\usepackage{bm}

\usepackage{float}
\usepackage[utf8]{inputenc}
\usepackage[T1]{fontenc}
\usepackage{mathptmx}
\usepackage{bbold}
\usepackage{siunitx}

\newcommand{\transformed}[1]{{#1}'}

\begin{document}


\title[Measurement circuit effects in three-terminal electrical transport measurements]{Measurement circuit effects in three-terminal electrical transport measurements}


\author{E. A. Martinez}
\altaffiliation[Author to whom correspondence should be addressed: \\ esmartin@microsoft.com]{}
 \affiliation{Microsoft Quantum Lab Copenhagen, Universitetsparken 5, 2100 Copenhagen, Denmark}

\author{A. Pöschl}
 \affiliation{Center for Quantum Devices, Niels Bohr Institute,
University of Copenhagen, Universitetsparken 5, 2100 Copenhagen, Denmark}

\author{E. B. Hansen}
 \affiliation{Microsoft Quantum Lab Copenhagen, Universitetsparken 5, 2100 Copenhagen, Denmark}

\author{M. A. Y. van de Poll}
 \affiliation{Center for Quantum Devices, Niels Bohr Institute,
University of Copenhagen, Universitetsparken 5, 2100 Copenhagen, Denmark}

\author{S. Vaitiekenas}
 \affiliation{Microsoft Quantum Lab Copenhagen, Universitetsparken 5, 2100 Copenhagen, Denmark}

\author{A. P. Higginbotham}%
 \affiliation{
Institute of Science and Technology Austria, 3400 Klosterneuburg, Austria
}

\author{L. Casparis}
 \affiliation{Microsoft Quantum Lab Copenhagen, Universitetsparken 5, 2100 Copenhagen, Denmark}

\date{\today}

\begin{abstract}
In this article we study measurement circuit effects in three-terminal electrical transport measurements arising from finite line impedances. We provide exact expressions relating the measured voltages and differential conductances to their values at the device under test, which allow for spurious voltage divider effects to be corrected. Finally, we test the implementation of these corrections with experimental measurements.
\end{abstract}

\maketitle

\section{Measurement circuit effects}
\label{sec:effects}

Three-terminal devices are a powerful tool for studying electrical transport in nanoscale systems. Three-terminal devices have been used to split Cooper pairs\cite{Hofstetter2009, Herrmann2010, Schindele2012, Michalek2015}, characterize end-to-end state correlations in nanowires\cite{Anselmetti2019}, study the local charge character of Andreev bound states \cite{Danon2020, PhysRevLett.124.036802, Yu2021} and have been proposed as probes for topological bulk properties\cite{Rosdahl2018, Pan2020, puglia2020closing}. In cryogenic experimental setups for electrical transport measurements, it is common for the measurement circuit to include filters, which give rise to finite line impedances. Such line impedances may give rise to spurious effects arising from voltage divider effects, particularly when studying the nonlocal conductance (cross-conductance) between terminals. These effects have been previously noted in the literature, and sometimes corrected to first order, see for example Ref.~\cite{Hofstetter2009}. The goal of this work is to give exact expressions relating the measured electrical quantities to the corresponding quantities at the device under test. These expressions allow to correct for voltage divider effects even in the regime where the device impedances are comparable to the line impedances, as long as the latter are well known.

We consider a device under test with three terminals: left ($l$), right ($r$) and ground ($g$). The device terminals are connected through line impedances $Z_l$, $Z_r$ and $Z_g$ to external measurement terminals and an external reference ground, respectively. The line impedances are part of the measurement circuit, and are assumed to be constant and linear. At the measurement terminals we apply voltages $V_l$, $V_r$ with respect to the external reference ground voltage $V_g$. The voltages at the (internal) device terminals are denoted $U_l$, $U_r$ and $U_g$ respectively. All the voltages may in general be time-dependent, and are assumed to have a DC component for biasing the device and AC components for lock-in differential conductance measurements. Typically the line impedances originate from low-pass filters, which also have capacitive shunts to ground. The results in this paper hold for signal frequencies much lower than the filter cut-off frequency, since above this frequency the setup effectively has more than three terminals.
\begin{figure}[ht]
    \centering
    \includegraphics[width=\columnwidth]{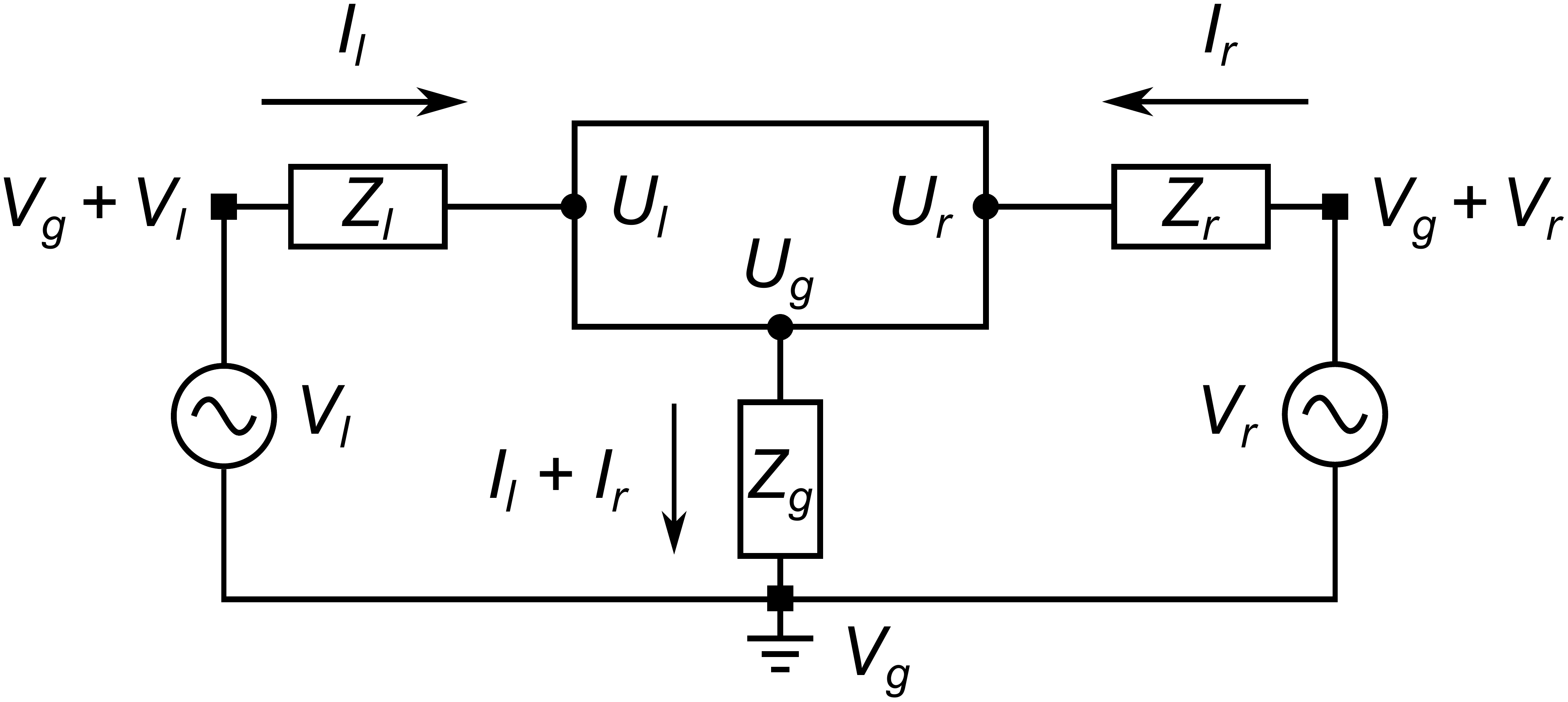}
    \caption{\label{fig:circuit}Measurement circuit for a generic three-terminal device. DC + AC voltages $V_l$ and $V_r$ are applied at the measurement terminals (squares) with respect to the external reference ground voltage $V_g$. The left/right measurement terminals and the external reference ground are respectively connected through line impedances $Z_l$, $Z_r$ and $Z_g$ to the device terminals. The voltages at the left, right and ground device terminals (circles) are $U_l$, $U_r$ and $U_g$ respectively.}
\end{figure}

The voltages $U_i$ at the device terminals are related to the voltages $V_i$ applied at the measurement terminals by:
\begin{align}
    U_l &= V_g + V_l - I_l Z_l \\
    U_r &= V_g + V_r - I_r Z_r \\
    U_g &= V_g + (I_l + I_r) Z_g.
\end{align}

Since only relative voltages are physically meaningful, we can define as the voltage reference the ground terminal at the device instead of the external measurement ground, that is, $U_g = 0$. Combining the previous expressions and eliminating $V_g$ we get:
\begin{align}
    U_l &= V_l - I_l Z_l - (I_l + I_r) Z_g \\
    U_r &= V_r - I_r Z_r - (I_l + I_r) Z_g,
\end{align}
or in vector form, where $\mathbf{V} = (V_l, V_r)^T$ and likewise for $\mathbf{U}$ and $\mathbf{I}$:
\begin{align}
    \label{eq:v_tilde}
    \mathbf{U} &= \mathbf{V} - Z \ \mathbf{I},
\end{align}
where $Z$ is a matrix with the line impedances:
\begin{align}
    Z = \left(
    \begin{matrix}
        Z_l + Z_g & Z_g\\
        Z_g & Z_r + Z_g
    \end{matrix}
    \right).
\end{align}
Eq.~(\ref{eq:v_tilde}) is particularly useful to calculate the DC voltage biases ${\mathbf{U}}$ at the device from the DC voltages $\mathbf{V}$ applied at the measurement terminals and the measured DC currents $\mathbf{I}$, by taking the zero-frequency component.

The electrical behavior of any device in a particular configuration is fully characterized by its differential conductance matrix as a function of the voltages at the terminals. The conductance matrix $\transformed{G}$ at the device is given by:
\begin{align}
    \label{eq:g_tilde}
    \transformed{G}(\mathbf{U}) &= \frac{\partial \mathbf{I}}{\partial \mathbf{U}} \nonumber \\
    &= \left(
    \begin{matrix}
        \frac{\partial I_l}{\partial U_l} & \frac{\partial I_l}{\partial U_r} \\
        \frac{\partial I_r}{\partial U_l} & \frac{\partial I_r}{\partial U_r} \\
    \end{matrix} \right).
\end{align}
Each of the elements of this matrix is a function of the voltages $\mathbf{U}$ at the device under test. Only these $2 \times 2$ elements are required to completely define the conductance matrix of the device, since the rest of its elements are determined by conservation of current and the choice of reference voltage.

In order to optimize signal-to-noise ratio it is usual to measure differential conductance using low-frequency lock-in techniques. AC voltage excitations $d\mathbf{V}$ with different frequencies are applied at the left and right measurement terminals, and the resulting AC currents $d\mathbf{I}$ are measured on both terminals at both frequencies, as explained in ref.~\cite{Anselmetti2019}. In order to directly measure the conductance matrix $\transformed{G}$, we would need to measure as well the AC voltage amplitudes $d\mathbf{U}$ at the device terminals. However, this is often not experimentally convenient for three-terminal devices, since three additional voltage probes and lock-in amplifiers are required. We can instead measure the conductance matrix $G$ with respect to the AC voltage excitations applied at the external measurement terminals, which is defined as:
\begin{align}
    G(\mathbf{V}) &= \frac{\partial \mathbf{I}}{\partial \mathbf{V}} \nonumber \\
    &= \left(
    \begin{matrix}
        \frac{\partial I_l}{\partial V_l} & \frac{\partial I_l}{\partial V_r} \\
        \frac{\partial I_r}{\partial V_l} & \frac{\partial I_r}{\partial V_r} \\
    \end{matrix} \right).
\end{align}
Both conductance matrices $G$ and $\transformed{G}$ agree for zero line impedances. But in the typical case of finite line impedances, the measured conductance matrix $G$ may show spurious effects that do not relate to the behavior of the device under test. 

We now want to convert the measured conductance matrix $G$ to the matrix $\transformed{G}$ at the device. Applying the chain rule on (\ref{eq:g_tilde}):
\begin{align}
    \transformed{G}(\mathbf{U}) &= \left(
        \begin{matrix}
            \frac{\partial I_l}{\partial V_l} & \frac{\partial I_l}{\partial V_r} \\
            \frac{\partial I_r}{\partial V_l} & \frac{\partial I_r}{\partial V_r} \\
        \end{matrix} \right) \left(
        \begin{matrix}
            \frac{\partial V_l}{\partial U_l} & \frac{\partial V_l}{\partial U_r} \\
            \frac{\partial V_r}{\partial U_l} & \frac{\partial V_r}{\partial U_r} \\
        \end{matrix} \right) \\
        \label{eq:g_tilde_2}
        &= G(\mathbf{V}) \left( \frac{\partial \mathbf{V}}{\partial \mathbf{U}} \right).
\end{align}
The conductance matrix $G'$ at the device is evaluated at the voltages $\mathbf{U}$ corresponding to the applied voltages $\mathbf{V}$ through Eq.~(\ref{eq:v_tilde}). Note that, in general, setting one of the $V_i$'s to be constant and varying the other will vary both $U_l$ and $U_r$. If one seeks to measure the conductance matrix at particular values of $\mathbf{U}$, the applied voltages $\mathbf{V}$ have to be chosen using Eq.~(\ref{eq:v_tilde}) such that at each measurement point $\mathbf{U}$ has the desired value.

The Jacobian matrix $\partial \mathbf{V} / \partial \mathbf{U}$ in Eq.~(\ref{eq:g_tilde_2}) relates the voltages outside of the measurement setup to the voltages at the device. We can calculate it by differentiating (\ref{eq:v_tilde}) with respect to $\mathbf{V}$:
\begin{align}
    \frac{\partial \mathbf{U}}{\partial \mathbf{V}} &= \mathbb{1} - Z \ \frac{\partial \mathbf{I}}{\partial \mathbf{V}}  \nonumber \\
    &= \mathbb{1} - Z \ G(\mathbf{V}).
\end{align}
Finally, inverting the previous expression using the inverse Jacobian rule:
\begin{align}
    \label{eq:inverse-jacobian}
    \frac{\partial \mathbf{V}}{\partial \mathbf{U}} &= ( \mathbb{1} - Z \ G(\mathbf{V}) )^{-1}.
\end{align}
Plugging (\ref{eq:inverse-jacobian}) into (\ref{eq:g_tilde_2}) we obtain the main result of this paper, the transformation from the raw conductance matrix $G(\mathbf{V})$ at the measurement terminals to the corrected conductance matrix $\transformed{G}(\mathbf{U})$ at the device:
\begin{align}
    \label{eq:g-transform}
    \transformed{G}(\mathbf{U}) &= G(\mathbf{V}) ( \mathbb{1} - Z \ G(\mathbf{V}) )^{-1}.
\end{align}

Note that the transformation in Eq.~(\ref{eq:g-transform}) is exact, and holds even when all the measured conductances are large compared to the line resistances. This is experimentally relevant since it is often convenient to measure a device in a highly conductive regime for signal-to-noise purposes. In particular, the expression in Eq.~(\ref{eq:g-transform}) is more general than the first-order voltage divider effects that are sometimes corrected for in the literature, e.g.~\cite{Hofstetter2009}, which hold when the conductances $G(\mathbf{V})$ are small compared to the line resistances $Z$. In order to derive the first-order corrections from the exact transformation we can expand Eq.~(\ref{eq:g-transform}) as a function of the product $Z \ G(\mathbf{V})$. The corrections are small if all the elements $(Z \ G(\mathbf{V}))_{ij} \ll 1$, that is:
\begin{align}
    |Z_l| &\ll |G_{li}|^{-1} \nonumber\\
    |Z_r| &\ll |G_{ri}|^{-1} \nonumber\\
    |Z_g| &\ll |G_{ij}|^{-1}, \nonumber
\end{align}
for $i, j = {l, r}$. If this is the case, we can expand the factor $( \mathbb{1} - Z \ G(\mathbf{V}) )^{-1}$ in Eq.~(\ref{eq:g-transform}) as a geometric series, obtaining to first order:
\begin{align}
    \label{eq:transform-first-order}
    \transformed{G}(\mathbf{U}) &\approx G(\mathbf{V}) ( \mathbb{1} + Z \ G(\mathbf{V}) ).
\end{align}
If the nonlocal conductances are much smaller than the local conductances, i.e. $|G_{lr}|, |G_{rl}| \ll |G_{ll}|, |G_{rr}|$, to zeroth order the correction can be written as:
\begin{align}
    \label{eq:correction-approx}
    \transformed{G}(\mathbf{U}) &\approx G(\mathbf{V}) + \left(
        \begin{matrix}
            (Z_l + Z_g) G_{ll}^2 & Z_g G_{ll} G_{rr} \\
            Z_g G_{ll} G_{rr} & (Z_r + Z_g) G_{rr}^2 \\
        \end{matrix} \right).
\end{align}

Assuming that the local conductances are real and positive, the local conductances $G'_{ll}, G'_{rr}$ at the device are larger than the measured ones by a correction quadratic in the conductance and proportional to the line impedance. This correction arises from the drop of the excitation voltages on the line impedances. Meanwhile, the nonlocal conductances $G'_{lr}, G'_{rl}$ at the device get a positive correction term proportional to the ground line impedance and both local conductances. This correction can be interpreted as a voltage divider effect: if the ground line impedance is finite, at the device ground there is a residual voltage excitation proportional to the conductance of one side, which then drops over the opposite side. This is the first-order effect corrected for in ref. \cite{Hofstetter2009}. Note that this voltage divider effect is an attenuated mirror image of the local conductances. This has an important consequence for experiments: if a nonlocal conductance signature is not proportional to the local conductances, then it cannot be attributed to voltage divider effects.

\section{Experimental validation}

\begin{figure}[ht]
    \centering
    \includegraphics[width=\columnwidth]{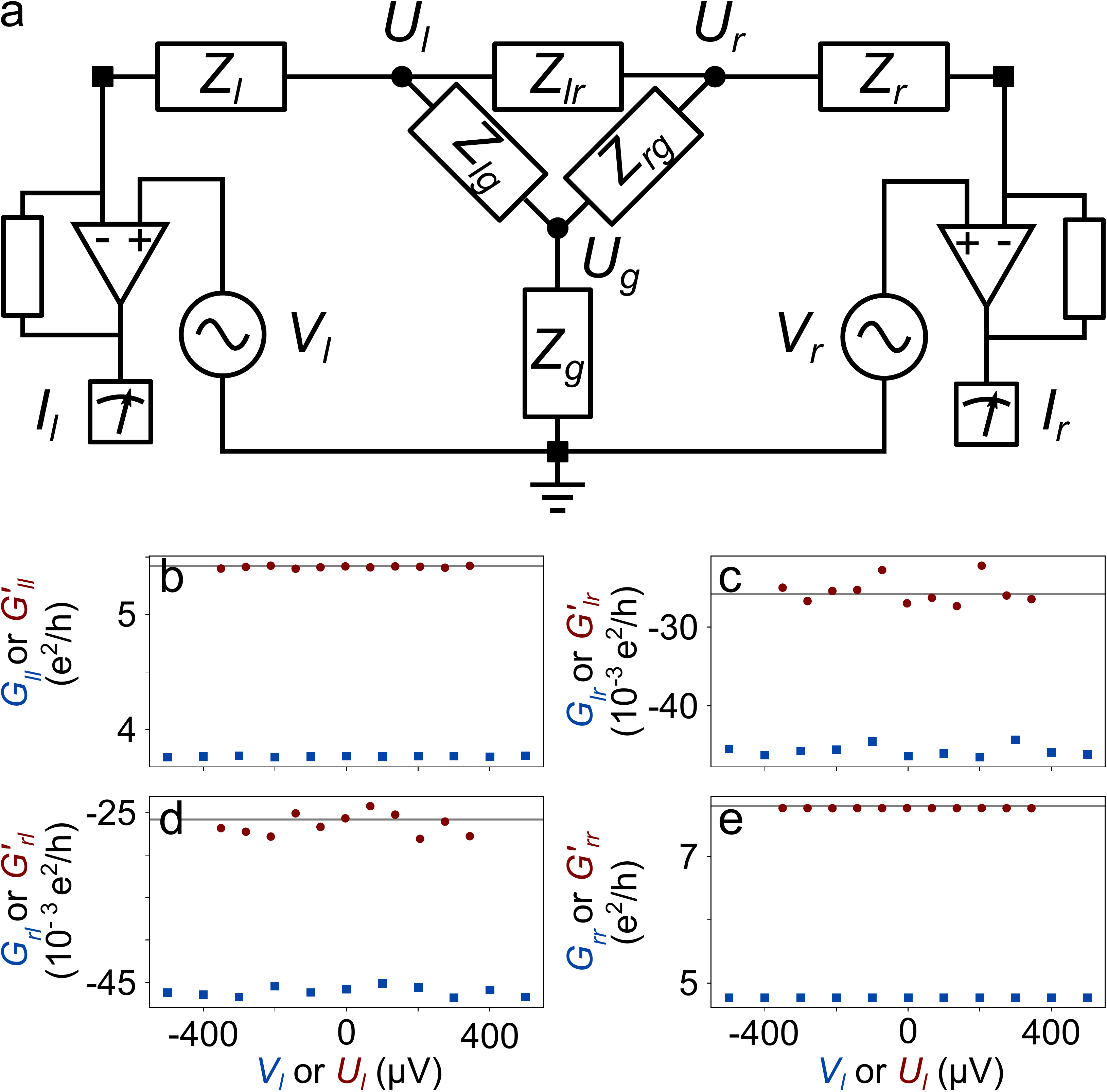}
    \caption{\label{fig:test_measurements}\textbf{a} The device under test is a resistor network at room temperature with $Z_{lr} = \SI{1}{\Mohm}$, $Z_{lg} = \SI{4760}{\ohm}$, $Z_{rg} = \SI{3313}{\ohm}$. The total line resistances are $Z_l = Z_r = \SI{2036}{\ohm}$, $Z_g = \SI{50}{\ohm}$. \textbf{b, c, d, e} Conductance matrix measurements as a function of the left bias voltage. The solid grey lines show the nominal conductance values of the test resistor network. The blue squares (lower) are the raw measured values of the conductance matrix elements $G_{ij}$ as a function of the voltage $V_l$ at the measurement terminals. The red circles (upper) are the corrected values of the conductance matrix $G'_{ij}$ as a function of the voltage $U_l$ at the device.}
\end{figure}

To validate experimentally the implementation of the measurement circuit effect corrections, we measured a three-terminal test resistor network using the measurement circuit depicted in Fig.~\ref{fig:test_measurements}a. The currents are measured using a Basel 983 I/V converter on each of the left and right measurement terminals. AC voltage excitations at $\SI{15}{\hertz}$ and $\SI{25}{\hertz}$ are applied at the voltage bias terminals of the left and right I/V converters, respectively. The excitation frequencies were chosen to be much lower than the I/V converter low-pass cutoff frequency of $\SI{1}{\kilo\hertz}$ to avoid finite frequency effects. The I/V converters have an input impedance of $\SI{33}{\ohm}$ that contributes to the total line resistances.

Fig.~\ref{fig:test_measurements}b-e show both the raw values of the measured conductance matrix $G(\mathbf{V})$ and the transformed conductance matrix $\transformed{G}(\mathbf{U})$. The first effect of the correction is to rescale the voltage axis to account for voltage drops over the line resistances, which result in a reduced DC voltage bias at the device. As discussed in the previous section, the measured local conductances in Fig.~\ref{fig:test_measurements}b,e (blue dots) are smaller than the conductances at the device, since the AC voltage excitations at the device are smaller than those applied at the measurement terminals. The corrected local conductances (red dots) agree with those expected from the nominal $Z_{lg}$ and $Z_{rg}$ resistor values (solid grey lines). Finally, the raw measured nonlocal conductances in Fig.~\ref{fig:test_measurements}c,d (blue dots) have a spurious contribution arising from voltage divider effects. The finite ground resistance gives rise to a finite AC voltage excitation at the device ground terminal when exciting from either of the sides. The corrected nonlocal conductance (red dots) takes these effects into account, yielding values in good agreement with the nominal $Z_{lr}$ resistor value (solid grey lines).

\begin{figure}[ht]
    \centering
    \includegraphics[width=\columnwidth]{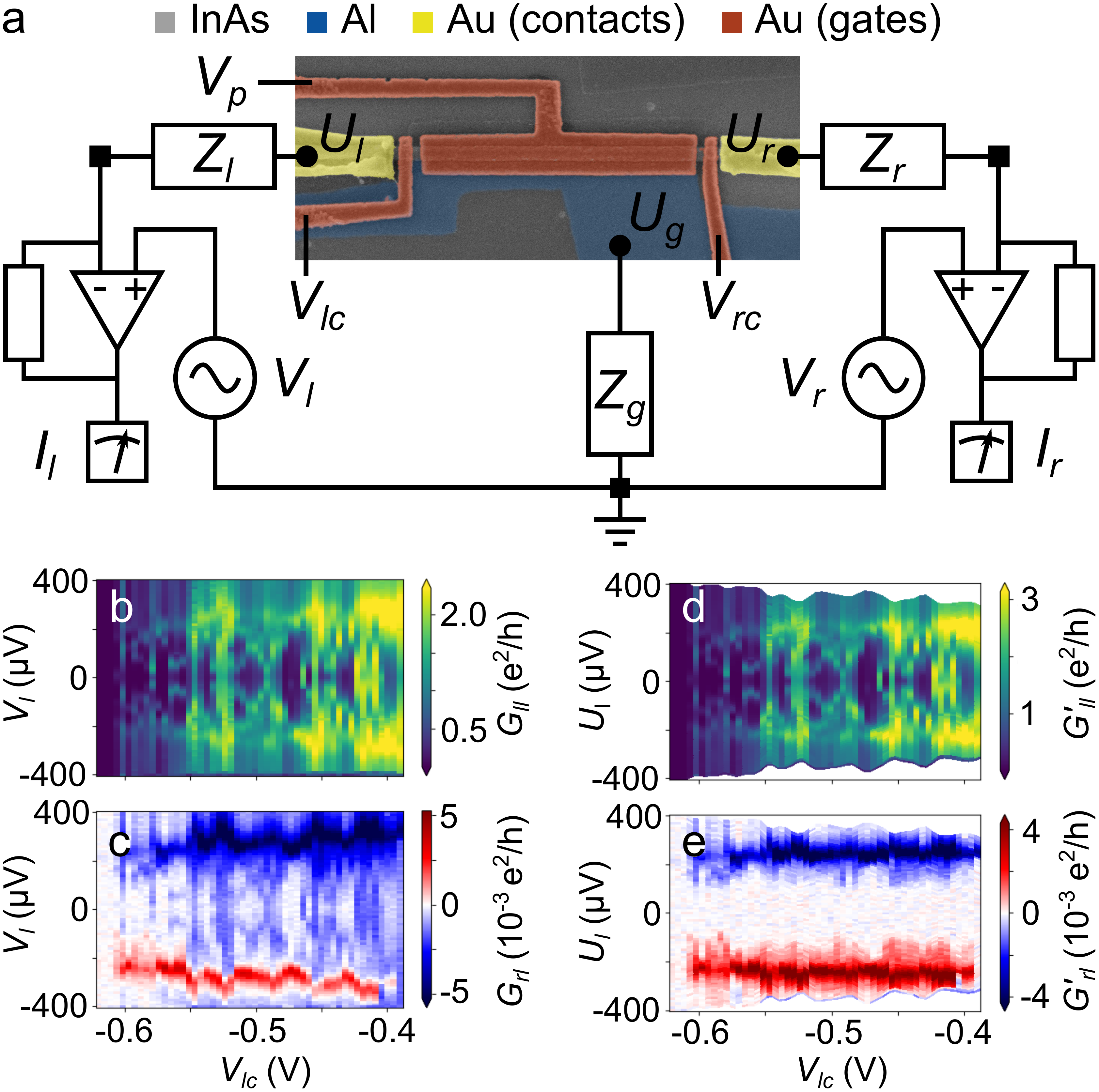}
    \caption{\label{fig:exp_corrected}\textbf{a} Three-terminal device micrograph and measurement circuit. \textbf{b}, \textbf{c}, Raw measured local (\textbf{b}) and nonlocal (\textbf{c}) conductances $G(\mathbf{V})$ when exciting the left measurement terminal, as a function of the applied DC voltage bias $V_l$ and left cutter gate voltage $V_{\text{lc}}$. \textbf{d}, \textbf{e}, Respective measurements corrected for measurement circuit effects. The vertical voltage axis has been transformed to the DC voltage $U_l$ at the left device terminal by Eq.~(\ref{eq:v_tilde}). The conductances $\transformed{G}(\mathbf{U})$ are now calculated with respect to the AC voltage excitation $dU_l$ at the left device terminal using Eq.~(\ref{eq:g-transform}).}
\end{figure}

In Fig.~\ref{fig:exp_corrected} we show conductance measurements at 20 mK on a hybrid superconducting-semiconducting three-terminal device (Fig.~\ref{fig:exp_corrected}a). An InAs nanowire (light grey, center) is contacted on its lower facet by an epitaxially-grown Al lead (blue). Gold ohmic contacts (yellow) are deposited on both ends of the nanowire, and electrostatic gates (red) are deposited on top: left/right cutters (lc/rc) and plunger (p). The line resistances are $Z_l = Z_r = 1830 \ \Omega$ and $Z_g = 915 \ \Omega$ and primarily originate from low-pass filters in the cryostat DC lines. The raw measured data (Figs.~\ref{fig:exp_corrected}b,c) shows superconducting coherence peaks, typical for tunneling spectroscopy of these devices, whose energy fluctuates as a function of the cutter gate voltage, moving outwards whenever the conductance $G_{ll}$ increases \cite{Chang2015}. In Fig.~\ref{fig:exp_corrected}c the raw nonlocal conductance $G_{rl}$ shows voltage divider artifacts in the form of a suppressed mirror image of the local conductance $G_{ll}$, as discussed in Section \ref{sec:effects}. Note  that, like the local conductance $G_{ll}$, the voltage divider effects are symmetric as a function of bias voltage. In the corrected datasets (Fig.~\ref{fig:exp_corrected}d,e) the energy of the superconducting coherence peaks is nearly constant as a function of cutter gate voltage. As shown in Fig.~\ref{fig:exp_corrected}e, the corrected nonlocal conductance at the device $\transformed{G}_{rl}$ no longer shows voltage divider artifacts, and is predominantly anti-symmetric as a function of bias voltage.

\begin{figure}[ht]
    \centering
    \includegraphics[width=\columnwidth]{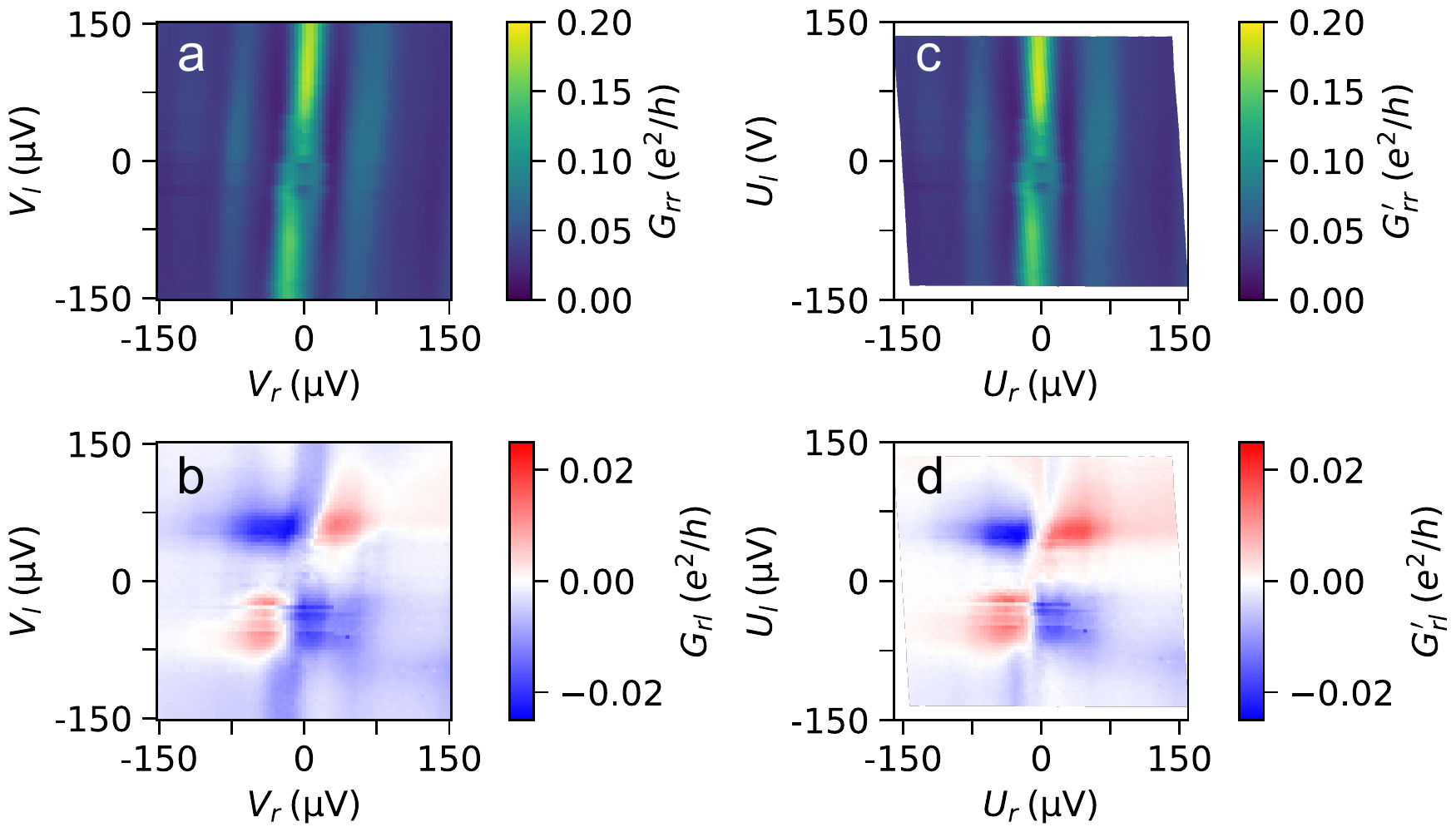}
    \caption{\label{fig:bias_bias} $\mathbf{a}$ and $\mathbf{b}$ depict two elements of the uncorrected conductance matrix $G$ with respect to the externally applied DC voltages $\mathbf{V}$. $\mathbf{c}$ and $\mathbf{d}$ show the same elements of the conductance matrix as $\mathbf{a}$ and $\mathbf{b}$, but after correction for voltage divider effects as a function of the DC voltages $\mathbf{U}$ at the device terminals. }
\end{figure}

In order to illustrate the dependence of $G$ and $\transformed{G}$ on $\mathbf{U}$ and $\mathbf{V}$, we depict a measurement of $G_{rr}(\mathbf{V})$ and $G_{rl}(\mathbf{V})$ together with the transformed matrix elements $\transformed{G}_{rr}(\mathbf{U})$ and $\transformed{G}_{rl}(\mathbf{U})$ as a function of both left and right bias in Fig.~\ref{fig:bias_bias}. The measurement is from a superconductor-semiconductor hybrid device identical to the one in Fig.~\ref{fig:circuit}. $G_{rr}$ and $G_{rl}$ depend on both $V_l$ and $V_r$ as a result of voltage divider effects. The nonlocal conductance $G_{rl}$ shows artifacts stemming from voltage divider effects in the form of negative conductance, clearly visible for values of $\mathbf{V}$ where $G_{rr}$ reaches values above $0.1\,e^2/h$. Those artifacts are absent in $\transformed{G}_{rl}$, which is obtained after transforming according to equation (\ref{eq:transform-first-order}). $\transformed{G}_{rr}$ and $\transformed{G}_{rl}$ in Fig.~\ref{fig:bias_bias} are plotted with respect to the voltages at the device terminals $\mathbf{U}$, which are calculated according to Eq.~(\ref{eq:v_tilde}). While the local conductance $G_{rr}$ has a slope as a function of both applied voltages $V_l$ and $V_r$, the transformed $\transformed{G}_{rr}$ is straight and depends mainly on the transformed voltage $U_r$. The nonlocal conductance $\transformed{G}_{rl}$ depends on both left and right bias coordinates, even after transforming to the voltages at the device.

In summary, we have derived exact expressions (Eqs. (\ref{eq:v_tilde}) and (\ref{eq:g-transform})) for correcting measurement circuit effects arising from finite line impedances in three-terminal electrical measurements, and validated them experimentally. These corrections, to first order, re-scale voltage biases and local conductance to account for voltage drops on the line impedances, as well as eliminate artifacts in the nonlocal conductance arising from voltage divider effects. This makes it possible to measure three-terminal devices in larger conductance regimes than otherwise possible, which is important for maximizing the signal-to-noise ratio of the nonlocal conductance.

\begin{acknowledgments}
We would like to acknowledge the Purdue selective area growth team Sergei Gronin, Ray Kallaher, Geoff Gardner and Michael Manfra for providing us with material, as well as Maren Kloster and Shivendra Upadhyay for fabricating the three-terminal device. We further would like to thank Denise Puglia for constructive discussions.
\end{acknowledgments}

\bibliography{main}

\end{document}